# Background field technique and renormalization in lattice gauge theory

Martin Lüscher

Deutsches Elektronen-Synchrotron DESY
Notkestrasse 85, D-22603 Hamburg, Germany

Peter Weisz

Max-Planck-Institut für Physik
Föhringer Ring 6, D-80805 München, Germany

**Abstract**

Lattice gauge theory with a background gauge field is shown to be renormalizable to all orders of perturbation theory. No additional counterterms are required besides those already needed in the absence of the background field. The argument closely follows the treatment given earlier for the case of dimensional regularization by Kluberg-Stern and Zuber. It is based on the BRS, background gauge and shift symmetries of the lattice functional integral.

# 1. Introduction

The background field technique [1] has long proved to be a useful tool to study the renormalization of non-abelian gauge theories. In particular, using dimensional regularization, the extended symmetry properties of the functional integral in the presence of a background gauge field have been exploited to establish the renormalizability of such theories to all orders of perturbation theory [2]. Moreover it has been shown that the renormalization of the effective action (including the dependence on the background field) does not require any further counterterms besides those already needed in the absence of the background field.

In lattice gauge theory the background field technique has been employed to compute the relation between the bare lattice coupling and the $\overline{\rm MS}$ coupling in SU($N$) gauge theories to one-loop order [3,4] and more recently to two-loop order [5,6]. An assumption implicitly made in these calculations is that the renormalization theorem mentioned above carries over to the lattice theory. We here show that this is indeed the case, thus putting our two-loop calculations [5,6] on solid grounds.

The proof of the theorem is not as difficult as it may seem, because the renormalizability of lattice gauge theories without background fields has already been established by Reisz some time ago [7–9]. We only need to show that the background field does not affect the renormalization of the theory and that the resulting effective action is the same as the one computed with dimensional regularization (up to a field-independent renormalization).

In sect. 2 we set up the background field formalism using dimensional regularization. There is nothing new in this section, but to understand the basic argumentation it is helpful to first consider this case. We then discuss the symmetries and the renormalization of the background field effective action (sects. 3,4). The lattice theory is considered in sect. 5. After introducing the background field, we show that the lattice functional integral has all the symmetry properties required for the proof of renormalizability (sect. 6). The latter is discussed in sect. 7 and we finally address the question of universality in the concluding sect. 8. Our notational conventions are collected in appendix A.

## 2. Background field formalism

In this section we consider the pure $SU(N)$ gauge theory in $D = 4 - 2\epsilon$ dimensions. The theory is quantized in euclidean space through the functional integral. The inclusion of matter fields is trivial and will not be discussed. For unexplained notations see appendix A.

### 2.1 Classical action and gauge fixing

The Yang-Mills action of an $SU(N)$ gauge potential $A_\mu(x)$ may be written as

$$S[A] = -\frac{1}{2g_0^2} \int d^D x \; \text{tr}\{F_{\mu\nu} F_{\mu\nu}\}, \tag{2.1}$$

where $g_0$ denotes the bare gauge coupling and $F_{\mu\nu}(x)$ the field strength tensor. The introduction of a background field in this theory is tied up with the gauge fixing, i.e. we shall go from eq.(2.1) directly to the gauge fixed theory with background field.

A possible way to proceed is as follows. Let $B_\mu(x)$ be a smooth external gauge field, the background field, and define the "quantum" field $q_\mu(x)$ through

$$A_\mu = B_\mu + g_0 q_\mu. \tag{2.2}$$

For the gauge fixing term we take

$$S_{\text{gf}}[B, q] = -\lambda_0 \int d^D x \; \text{tr}\{D_\mu q_\mu D_\nu q_\nu\}, \tag{2.3}$$

where $\lambda_0$ is the bare gauge fixing parameter and

$$D_\mu = \partial_\mu + \text{Ad} B_\mu. \tag{2.4}$$

The action of the associated ghost fields $c$ and $\bar{c}$ reads

$$S_{\text{FP}}[B, q, \bar{c}, c] = -2 \int d^D x \; \text{tr}\{D_\mu \bar{c}\, (D_\mu + g_0 \text{Ad} q_\mu) c\}, \tag{2.5}$$

and the total action of the gauge fixed theory is given by

$$S_{\text{tot}}[B, q, \bar{c}, c] = S[B + g_0 q] + S_{\text{gf}}[B, q] + S_{\text{FP}}[B, q, \bar{c}, c]. \tag{2.6}$$

Note that the background field is not required to satisfy the Yang-Mills field equations. It is just an arbitrary external source field, which is coupled to the "quantum" fields $q$, $\bar{c}$ and $c$ in some particular way.

In the following we shall not refer to the derivation of the total action given above. One may hence adopt the point of view that the theory discussed in this paper is defined by eq.(2.6). Note that the standard action with the usual covariant gauge fixing term is recovered for vanishing background field.

*2.2 Generating functionals*

We now introduce classical source fields $J_\mu(x), \bar{\eta}(x)$ and $\eta(x)$ for the quantum fields and consider the partition function

$$Z[B, J, \bar{\eta}, \eta] = \frac{1}{\mathcal{N}} \int \mathrm{D}[q]\mathrm{D}[\bar{c}]\mathrm{D}[c]$$
$$\times \exp\left\{-S_{\mathrm{tot}}[B, q, \bar{c}, c] + (J, q) + (\bar{\eta}, c) + (\bar{c}, \eta)\right\}. \quad (2.7)$$

The normalization factor $\mathcal{N}$ is chosen so that $Z[0, 0, 0, 0] = 1$. Since the total action is a polynomial in the background field, it is straightforward to expand the partition function in powers of $B, J, \bar{\eta}$ and $\eta$. The coefficients in this expansion are expectation values of products of local operators evaluated at vanishing sources. In particular, the usual Feynman rules with dimensional regularization apply at this point.

In the following we shall always think of $Z[B, J, \bar{\eta}, \eta]$ in this way, namely as a formal power series in $B, J, \bar{\eta}$ and $\eta$. To any order of perturbation theory the functional is then completely well-defined.

If $\mathcal{O}$ is a polynomial in the fields $B, q, \bar{c}$ and $c$, its expectation value in the presence of the background field is given by

$$\langle \mathcal{O} \rangle_B = \frac{1}{\mathcal{N}_B} \int \mathrm{D}[q]\mathrm{D}[\bar{c}]\mathrm{D}[c]\, \mathcal{O}[B, q, \bar{c}, c] \exp\left\{-S_{\mathrm{tot}}[B, q, \bar{c}, c]\right\}. \quad (2.8)$$

The normalization factor is here fixed by the condition $\langle 1 \rangle_B = 1$. The remarks made above apply to $\langle \mathcal{O} \rangle_B$ as well, i.e. after expanding in powers of the background field, such expectation values are well-defined and computable in perturbation theory.

In momentum space the expansion of the free energy

$$W[B, J, \bar{\eta}, \eta] = \ln\left(Z[B, J, \bar{\eta}, \eta]\right) \quad (2.9)$$

assumes the form

$$W[B, J, \bar{\eta}, \eta] = \sum_{j,k,l=0}^{\infty} \frac{1}{j!\,k!\,(l!)^2} \int \frac{\mathrm{d}^D p_1}{(2\pi)^D} \cdots \frac{\mathrm{d}^D s_l}{(2\pi)^D}$$

$$\times (2\pi)^D \delta(p_1 + \ldots + s_l)\, G^{(j,k,l)}(p_1, \ldots, s_l)^{a_1 \ldots d_l}_{\mu_1 \ldots \nu_k}$$

$$\times \tilde{B}^{a_1}_{\mu_1}(-p_1) \ldots \tilde{B}^{a_j}_{\mu_j}(-p_j)\, \tilde{J}^{b_1}_{\nu_1}(-q_1) \ldots \tilde{J}^{b_k}_{\nu_k}(-q_k)$$

$$\times \tilde{\bar{\eta}}^{c_1}(-r_1) \ldots \tilde{\bar{\eta}}^{c_l}(-r_l)\, \tilde{\eta}^{d_1}(-s_1) \ldots \tilde{\eta}^{d_l}(-s_l). \quad (2.10)$$

The coefficient functions $G^{(j,k,l)}$ appearing here are just the connected parts of the corresponding coefficient functions which occur in the analogous expansion of the partition function. In particular, to any order of perturbation theory they are sums of connected Feynman diagrams with the appropriate number and type of external lines (cf. ref.[6] for further details).

The effective action (or vertex functional) of the theory is finally obtained by a Legendre transformation. Explicitly, if we introduce the fields

$$Q^a_\mu(x) = \frac{\delta W}{\delta J^a_\mu(x)}, \quad (2.11)$$

$$C^a(x) = \frac{\delta W}{\delta \bar{\eta}^a(x)}, \quad (2.12)$$

$$\overline{C}^a(x) = -\frac{\delta W}{\delta \eta^a(x)}, \quad (2.13)$$

the effective action is given by

$$\Gamma[B, Q, \overline{C}, C] = W[B, J, \bar{\eta}, \eta] - (J, Q) - (\bar{\eta}, C) - (\overline{C}, \eta). \quad (2.14)$$

The source fields $J, \bar{\eta}$ and $\eta$ on the right hand side of this equation are to be expressed as power series in $B, Q, \overline{C}$ and $C$ by inverting eqs.(2.11)–(2.13).

By expanding the effective action in powers of $B, Q, \overline{C}$ and $C$ [in a way analogous to eq.(2.10)] one obtains the vertex functions $\Gamma^{(j,k,l)}$. They are closely related to the correlation functions $G^{(j,k,l)}$. In particular, $\Gamma^{(0,2,0)}$ is equal to minus the inverse propagator of the quantum field,

$$\Gamma^{(0,2,0)}(q, -q)^{ab}_{\mu\nu} G^{(0,2,0)}(q, -q)^{bc}_{\nu\rho} = -\delta^{ac}\delta_{\mu\rho}, \quad (2.15)$$

and for the propagator of the ghost fields one finds

$$\Gamma^{(0,0,1)}(r,-r)^{ab}G^{(0,0,1)}(r,-r)^{bc} = -\delta^{ac}. \tag{2.16}$$

In all other cases the vertex function $\Gamma^{(j,k,l)}$ coincides with the one-particle irreducible full propagator amputated part of $G^{(j,k,l)}$. In the case of the background field propagator this implies

$$G^{(2,0,0)}(p,-p)^{ab}_{\mu\nu} = \Gamma^{(2,0,0)}(p,-p)^{ab}_{\mu\nu}$$
$$+ \Gamma^{(1,1,0)}(p,-p)^{aa'}_{\mu\mu'}\Gamma^{(1,1,0)}(-p,p)^{bb'}_{\nu\nu'}G^{(0,2,0)}(p,-p)^{a'b'}_{\mu'\nu'}, \tag{2.17}$$

and similar relations hold for the other correlation functions.

## 3. Symmetries

The theory formulated in sect. 2 has various symmetries which play an important rôle in the following since they are preserved by the renormalization procedure and thus restrict the possible form of the counterterms.

### 3.1 BRS transformations [10–13]

Suppose $F$ is some functional depending on the quantum fields $q, \bar{c}$ and $c$. We then define its BRS variation $\delta_{\text{BRS}} F$ through

$$\delta_{\text{BRS}} F = \int d^D x \left\{ \delta_{\text{BRS}} q^a_\mu \frac{\delta F}{\delta q^a_\mu} + \delta_{\text{BRS}} \bar{c}^a \frac{\delta F}{\delta \bar{c}^a} + \delta_{\text{BRS}} c^a \frac{\delta F}{\delta c^a} \right\}, \tag{3.1}$$

where

$$\delta_{\text{BRS}} q_\mu = (D_\mu + g_0 \text{Ad} q_\mu) c, \tag{3.2}$$

$$\delta_{\text{BRS}} \bar{c} = \lambda_0 D_\mu q_\mu, \tag{3.3}$$

$$\delta_{\text{BRS}} c = -g_0 c^2 \tag{3.4}$$

[cf. eq.(2.4)]. Note that we do not include an anti-commuting infinitesimal parameter in the definition of $\delta_{\text{BRS}} F$ (as is often done). We simply consider the BRS variation to be a linear mapping from the space of all functionals $F$ into itself. In particular, we can apply $\delta_{\mathrm{BRS}}$ more than once.

The BRS variation involves the background field through the derivatives $D_\mu$ in eqs.(3.2) and (3.3). This dependence is such that the identity

$$\delta_{\mathrm{BRS}} S_{\mathrm{tot}} = 0 \tag{3.5}$$

holds for all background fields. Moreover the a priori measure $\mathrm{D}[q]\mathrm{D}[\bar{c}]\mathrm{D}[c]$ is also left invariant by the transformation, which is, therefore, a true global symmetry of the system.

An immediate consequence of these observations is that

$$\langle \delta_{\mathrm{BRS}} F \rangle_B = 0 \tag{3.6}$$

for any functional $F$. This leads to a number of interesting identities among the correlation functions of the quantum fields. For example, if we take $F = \bar{c}^a(x)(D_\nu q_\nu)^b(y)$ and use the ghost field equation, we obtain

$$\langle (D_\mu q_\mu)^a(x)(D_\nu q_\nu)^b(y) \rangle_B = \lambda_0^{-1} \delta^{ab} \delta(x-y). \tag{3.7}$$

The longitudinal part of the propagator of the quantum field $q$ is hence fixed to its value at tree-level of perturbation theory. It should be emphasized that eq.(3.7) is an exact relation which holds for all background fields $B$.

### 3.2 Background gauge invariance

Let $\Lambda(x)$ be a (classical) gauge transformation. We define its action on the background field and the quantum fields through

$$B_\mu \to B_\mu^\Lambda = \Lambda B_\mu \Lambda^{-1} + \Lambda \partial_\mu \Lambda^{-1}, \tag{3.8}$$

$$q_\mu \to q_\mu^\Lambda = \Lambda q_\mu \Lambda^{-1}, \tag{3.9}$$

$$\bar{c} \to \bar{c}^\Lambda = \Lambda \bar{c} \Lambda^{-1}, \tag{3.10}$$

$$c \to c^\Lambda = \Lambda c \Lambda^{-1}. \tag{3.11}$$

It is evident that the total action $S_{\mathrm{tot}}$ and the a priori measure are invariant under such transformations. If we transform the classical fields $Q, \overline{C}$ and $C$ in the same way as the corresponding quantum fields, it is easy to show that

$$\Gamma[B^\Lambda, Q^\Lambda, \overline{C}^\Lambda, C^\Lambda] = \Gamma[B, Q, \overline{C}, C]. \tag{3.12}$$

A simple consequence of this symmetry is the transversality of the background field propagator $\Gamma^{(2,0,0)}(p, -p)_{\mu\nu}^{ab}$.

*3.3 Shift symmetry*

For any functional $F[B, q]$ the shift symmetry variation is defined by

$$[\delta_{\mathrm{s}} F]_\mu^a = g_0 \frac{\delta F}{\delta B_\mu^a} - \frac{\delta F}{\delta q_\mu^a}. \tag{3.13}$$

Note that this is a local operation. Any functional depending on $B$ and $q$ through the combination $B + g_0 q$ is invariant under the shift symmetry. A less trivial observation is that

$$[\delta_{\mathrm{s}} S_{\mathrm{tot}}]_\mu^a = \delta_{\mathrm{BRS}} \left[ (D_\mu + g_0 \mathrm{Ad} q_\mu) \bar{c} \right]^a. \tag{3.14}$$

The total action is hence not invariant, but its variation is equal to the BRS variation of some other quantity. This may be interpreted as an expression of the fact that the theory is invariant under a shift variation up to a change of the gauge fixing function.

A simple consequence of eq.(3.14) is that $W[B, 0, 0, 0]$ is independent of the background field (and thus equal to zero). In order to show this we note that

$$g_0 \frac{\delta}{\delta B_\mu^a} W[B, 0, 0, 0] = -\langle [\delta_{\mathrm{s}} S_{\mathrm{tot}}]_\mu^a \rangle_B. \tag{3.15}$$

From eqs.(3.14) and (3.6) it is then immediate that the expectation value on the right hand side of this equation vanishes.

It follows from this that the correlation functions $G^{(j,0,0)}$ are equal to zero for all $j$. This does not imply that the corresponding vertex functions $\Gamma^{(j,0,0)}$ vanish, but since they are closely related to the correlation functions, a number of interesting identities result. From eq.(2.17), for example, we deduce that

$$\Gamma^{(2,0,0)}(p, -p)_{\mu\nu}^{ab} =$$
$$- \Gamma^{(1,1,0)}(p, -p)_{\mu\mu'}^{aa'} \Gamma^{(1,1,0)}(-p, p)_{\nu\nu'}^{bb'} G^{(0,2,0)}(p, -p)_{\mu'\nu'}^{a'b'}. \tag{3.16}$$

# 4. Renormalization

The renormalizability of the theory in the presence of a background field has been established many years ago by Kluberg-Stern and Zuber [2]. It is not our aim here to review their work. We shall instead take it for granted that the theory without background field is renormalizable and proceed to show that the renormalizability is preserved when the background field is turned on. This is not as difficult as the complete proof of renormalizability given in ref.[2]. The argument will be presented in such a way that it carries over to the lattice theory with only marginal modifications.

### 4.1 BRS identity

Under a BRS variation the quantum fields transform non-linearly. To work out the consequences of the BRS symmetry for the vertex functional the transformation must first be linearized. Since

$$\delta_{\text{BRS}}(\delta_{\text{BRS}} q_\mu) = \delta_{\text{BRS}}(\delta_{\text{BRS}} c) = 0 \tag{4.1}$$

it is obvious that this can be achieved by including $\delta_{\text{BRS}} q_\mu$ and $\delta_{\text{BRS}} c$ in the list of fields. Following ref.[13] we thus add further source terms,

$$(K, \delta_{\text{BRS}} q) - (L, \delta_{\text{BRS}} c), \tag{4.2}$$

in eq.(2.7), where $K_\mu^a(x)$ and $L^a(x)$ are new source fields with ghost numbers $-1$ and $-2$, respectively. The free energy $W$ and the vertex functional $\Gamma$ are defined as before [eqs.(2.9) and (2.11)–(2.14)]. The background gauge invariance discussed in subsect. 3.2 is preserved by the new source terms if $K$ and $L$ are transformed in the same way as $Q, \overline{C}$ and $C$.

It is now straightforward to show that the free energy satisfies

$$\int \mathrm{d}^D x \left\{ J_\mu^a \frac{\delta W}{\delta K_\mu^a} + \bar\eta^a \frac{\delta W}{\delta L^a} + \lambda_0 \eta^a D_\mu^{ab} \frac{\delta W}{\delta J_\mu^b} \right\} = 0 \tag{4.3}$$

as a consequence of the BRS symmetry of the functional integral. In terms of the vertex functional this identity reads

$$\int \mathrm{d}^D x \left\{ \frac{\delta \Gamma}{\delta Q_\mu^a} \frac{\delta \Gamma}{\delta K_\mu^a} - \frac{\delta \Gamma}{\delta C^a} \frac{\delta \Gamma}{\delta L^a} + \lambda_0 (D_\mu Q_\mu)^a \frac{\delta \Gamma}{\delta \overline{C}^a} \right\} = 0. \tag{4.4}$$

It should be emphasized that this relation holds exactly to all orders in the source fields $B, Q, \overline{C}, C, K, L$ and the bare coupling $g_0$.

*4.2 Renormalized vertex functional*

We now define a renormalized vertex functional

$$\Gamma_R[B, Q, \overline{C}, C, K, L] = \Gamma[B, Z_3^{1/2}Q, \widetilde{Z}_3^{1/2}\overline{C}, \widetilde{Z}_3^{1/2}C, \widetilde{Z}_3^{1/2}K, Z_3^{1/2}L] \qquad (4.5)$$

and express the bare coupling and gauge parameter through a renormalized coupling $g$ and a renormalized gauge parameter $\lambda$ according to

$$g_0 = \mu^\epsilon Z_1 Z_3^{-3/2} g, \qquad \mu: \text{normalization mass}, \qquad (4.6)$$

$$\lambda_0 = Z_3^{-1} \lambda. \qquad (4.7)$$

The renormalization constants $Z_1, Z_3$ and $\widetilde{Z}_3$ are formal power series in $g$ with coefficients depending on $\epsilon$ and $\lambda$. They are chosen in such a way that $\Gamma_R[0, Q, \overline{C}, C, K, L]$ is finite for $\epsilon \to 0$ to all orders of $g$. It is well-known (and shown in ref.[13], for example) that this possible. Our aim below will be to prove that $\Gamma_R$ remains finite for non-zero background fields $B$.

The renormalized vertex functions $\Gamma_R^{(j,k,l)}$ are obtained by expanding the vertex functional $\Gamma_R[B, Q, \overline{C}, C, 0, 0]$ in powers of the source fields. In terms of the bare vertex functions they are given by

$$\Gamma_R^{(j,k,l)} = Z_3^{k/2} \widetilde{Z}_3^{l} \Gamma^{(j,k,l)}. \qquad (4.8)$$

The renormalized correlation functions

$$G_R^{(j,k,l)} = Z_3^{-k/2} \widetilde{Z}_3^{-l} G^{(j,k,l)} \qquad (4.9)$$

are related to the renormalized vertex functions in exactly the same way as the corresponding bare functions are [cf. eqs.(2.15)–(2.17)].

It is also trivial to verify that the background gauge invariance is not affected by the renormalization. The BRS identity (4.4) translates to

$$\int d^D x \left\{ \frac{\delta \Gamma_R}{\delta Q_\mu^a} \frac{\delta \Gamma_R}{\delta K_\mu^a} - \frac{\delta \Gamma_R}{\delta C^a} \frac{\delta \Gamma_R}{\delta L^a} + \lambda (D_\mu Q_\mu)^a \frac{\delta \Gamma_R}{\delta \overline{C}^a} \right\} = 0. \qquad (4.10)$$

The important point to note is that the renormalization constants do not appear in this equation.

*4.3 Finiteness of the renormalized vertex functional at $D = 4$*

We now prove that $\Gamma_R[B, Q, \overline{C}, C, K, L]$ is finite in the limit $\epsilon \to 0$ to all orders of the renormalized coupling $g$. Let $\Gamma_{R,l}$ be the contribution to the renormalized vertex functional at $l$-loop order of perturbation theory so that

$$\Gamma_R = \sum_{l=0}^{\infty} \Gamma_{R,l}. \tag{4.11}$$

The lowest-order expression is

$$\Gamma_{R,0} = -S_{\text{tot}}[B, Q, \overline{C}, C] + (K, \tilde{\delta}_{\text{BRS}} Q) - (L, \tilde{\delta}_{\text{BRS}} C), \tag{4.12}$$

where the bare parameters in the action, $g_0$ and $\lambda_0$, are to be replaced by $\mu^\epsilon g$ and $\lambda$. The classical fields $\tilde{\delta}_{\text{BRS}} Q_\mu$ and $\tilde{\delta}_{\text{BRS}} C$ are defined by

$$\tilde{\delta}_{\text{BRS}} Q_\mu = (D_\mu + \mu^\epsilon g \mathrm{Ad} Q_\mu) C, \tag{4.13}$$

$$\tilde{\delta}_{\text{BRS}} C = -\mu^\epsilon g C^2 \tag{4.14}$$

(the operator $\tilde{\delta}_{\text{BRS}}$ acts on the source fields and should not be confused with the BRS variation $\delta_{\text{BRS}}$ defined in sect. 3).

The proof of finiteness of $\Gamma_R$ proceeds inductively, i.e. we assume that $\Gamma_{R,l}$ is finite for all $l < n$ and then show that $\Gamma_{R,n}$ is finite, too. $\Gamma_{R,n}$ may be written as a sum of a singular part $\Delta \Gamma_{R,n}$ and a finite remainder. The induction hypothesis implies that the divergences of the proper subdiagrams of any $n$–loop diagram are cancelled by the counterterms. Without loss we may thus assume that the singular part $\Delta \Gamma_{R,n}$ is of the form

$$\Delta \Gamma_{R,n} = \int \mathrm{d}^D x \, p(x), \tag{4.15}$$

where $p(x)$ is a polynomial in the source fields and their derivatives at the point $x$. The divergent terms which can appear have engineering dimension less than or equal to 4 and ghost number 0. Moreover they must be invariant under the space-time symmetries and the background gauge invariance discussed in subsect. 3.2. And since the renormalized vertex functional is finite for zero background field we can insist that $p = 0$ if $B = 0$.

Taking these remarks into account it is straightforward to show that

$$\Delta \Gamma_{R,n} = \int \mathrm{d}^D x \, \mathrm{tr} \, \{c_1 G_{\mu\nu} G_{\mu\nu} + c_2 G_{\mu\nu} D_\mu Q_\nu + c_3 G_{\mu\nu} Q_\mu Q_\nu\} \tag{4.16}$$

is a sufficiently general ansatz for the singular part. The coefficients $c_1, c_2, c_3$ appearing in this expression may be assumed to be polynomials in $1/\epsilon$ with no constant term. $G_{\mu\nu}$ denotes the field strength tensor associated with the background field.

The BRS identity (4.10) must be satisfied order by order in the loop expansion. At $n$–loop order the singular part $\Delta\Gamma_{R,n}$ contributes to some of the terms on the left hand side of eq.(4.10). Since there are no other divergent terms to this order and since the sum of all terms must vanish we conclude that

$$\int d^D x \, \frac{\delta \Delta\Gamma_{R,n}}{\delta Q_\mu^a} \frac{\delta \Gamma_{R,0}}{\delta K_\mu^a} = \text{finite}. \tag{4.17}$$

The general form (4.16) of the singular part has here been taken into account. After inserting eq.(4.12) into eq.(4.17) one obtains

$$\int d^D x \, [(D_\mu + \mu^\epsilon g \mathrm{Ad} Q_\mu)C]^a \, \frac{\delta \Delta\Gamma_{R,n}}{\delta Q_\mu^a} = \text{finite}, \tag{4.18}$$

and since this must be true for arbitrary fields $C$, it follows that

$$\left[ D_\mu^{ab} + \mu^\epsilon g (\mathrm{Ad} Q_\mu)^{ab} \right] \frac{\delta \Delta\Gamma_{R,n}}{\delta Q_\mu^b} = \text{finite}. \tag{4.19}$$

Expanding in powers of $Q$ we conclude from this that $c_2$ and $c_3$ must be finite and so are equal to zero.

We have thus shown that the singular part is of the form

$$\Delta\Gamma_{R,n} = c_1 \int d^D x \, \mathrm{tr}\{G_{\mu\nu} G_{\mu\nu}\}. \tag{4.20}$$

In particular, all vertex functions $\Gamma_{R,n}^{(j,k,l)}$ with $k \geq 1$ or $l \geq 1$ are finite. The background field propagator, on the other hand, may have a divergent part given by

$$\Delta\Gamma_{R,n}^{(2,0,0)}(p,-p)_{\mu\nu}^{ab} = -2c_1 \delta^{ab} \left( \delta_{\mu\nu} p^2 - p_\mu p_\nu \right). \tag{4.21}$$

We now recall the identity (3.16) which was derived from the shift symmetry of the functional integral. In terms of renormalized vertex functions it reads

$$\Gamma_R^{(2,0,0)}(p,-p)_{\mu\nu}^{ab} =$$
$$- \Gamma_R^{(1,1,0)}(p,-p)_{\mu\mu'}^{aa'} \Gamma_R^{(1,1,0)}(-p,p)_{\nu\nu'}^{bb'} G_R^{(0,2,0)}(p,-p)_{\mu'\nu'}^{a'b'}. \tag{4.22}$$

Since the right hand side of this equation is finite at $n$–loop order, we immediately conclude that $c_1 = 0$. This completes the proof of finiteness of the renormalized vertex functional.

## 5. Lattice theory

In this section we consider the pure $SU(N)$ gauge theory on a four-dimensional hypercubic lattice $\Lambda$ with spacing $a$. Our lattice notations are standard (cf. appendix A). By abuse of language many symbols that have already appeared in the preceding sections will be employed again to denote the analogous lattice quantities.

*5.1 Background field and gauge fixing on the lattice*

The lattice formulation of the $SU(N)$ gauge theory with background field is not unique. Different lattice actions may be chosen and the precise way in which the background field is introduced is arbitrary to some extent. The differences between the choices that one has should be irrelevant in the continuum limit, if the lattice theory has all the symmetries required for the proof of renormalizability.

The Wilson action of a lattice gauge field $U(x,\mu)$ is given by

$$S[U] = \frac{1}{g_0^2} \sum_{x \in \Lambda} \sum_{\mu,\nu} \text{Re tr} \left\{ 1 - P(x,\mu,\nu) \right\}, \tag{5.1}$$

where $P(x,\mu,\nu)$ denotes the plaquette field [eq.(A.14)]. For definiteness we here choose this form of the lattice action, although other actions would do just as well. As in the continuum theory the introduction of the background field goes along with the gauge fixing. The latter has been worked out in detail in ref.[14], for example, so that here we can be rather brief.

We begin by introducing the background field $B_\mu(x)$ and the "quantum" field $q_\mu(x)$ through

$$U(x,\mu) = e^{g_0 a q_\mu(x)} e^{a B_\mu(x)}. \tag{5.2}$$

The advantage of having two exponential factors rather than one is that the background gauge transformations assume a simple form, viz. †

$$B_\mu^\Lambda(x) = \frac{1}{a} \ln \left\{ \Lambda(x) e^{aB_\mu(x)} \Lambda(x + a\hat{\mu})^{-1} \right\}, \tag{5.3}$$

$$q_\mu^\Lambda(x) = \Lambda(x) q_\mu(x) \Lambda(x)^{-1}. \tag{5.4}$$

This transformation amounts to an ordinary gauge transformation on the total field $U(x, \mu)$ so that the action $S[U]$ is invariant.

For the gauge fixing term we take

$$S_{\rm gf}[B, q] = -\lambda_0 a^4 \sum_{x \in \Lambda} \text{tr} \left\{ D_\mu^* q_\mu(x) D_\nu^* q_\nu(x) \right\}. \tag{5.5}$$

The symbol $D_\mu^*$ in this expression denotes a background gauge covariant difference operator. There are actually two such operators, corresponding to forward and backward differences, which will be needed in the following. Explicitly they are given by

$$D_\mu f(x) = \frac{1}{a} \left\{ e^{aB_\mu(x)} f(x + a\hat{\mu}) e^{-aB_\mu(x)} - f(x) \right\}, \tag{5.6}$$

$$D_\mu^* f(x) = \frac{1}{a} \left\{ f(x) - e^{-aB_\mu(x - a\hat{\mu})} f(x - a\hat{\mu}) e^{aB_\mu(x - a\hat{\mu})} \right\}, \tag{5.7}$$

for any lattice function $f(x)$ with values in the Lie algebra of SU($N$). It is now not difficult to verify that $S_{\rm gf}[B, q]$ is invariant under background gauge transformations.

The ghost field action associated with this gauge fixing term reads [14]

$$S_{\rm FP}[B, q, \bar{c}, c] =$$
$$-2a^4 \sum_{x \in \Lambda} \text{tr} \left\{ D_\mu \bar{c}(x) \left[ J\left(g_0 a q_\mu(x)\right)^{-1} D_\mu + g_0 \text{Ad} q_\mu(x) \right] c(x) \right\}. \tag{5.8}$$

The matrix $J$ appearing here is the differential of the exponential mapping (cf. subsect. A.5). Under background gauge transformations the ghost fields $c$ and $\bar{c}$ transform in the same way as the quantum field $q$. $S_{\rm FP}[B,q,\bar{c},c]$ is invariant under such transformations and so is the total action

$$S_{\rm tot}[B,q,\bar{c},c] = S[U] + S_{\rm gf}[B,q] + S_{\rm FP}[B,q,\bar{c},c]. \tag{5.9}$$

---

† The background field $B_\mu(x)$ and the gauge transformation $\Lambda(x)$ are classical external fields. Without loss we may restrict attention to fields such that $a\|B_\mu(x)\|$ is small and $\Lambda(x)\Lambda(x + a\hat{\mu})^{-1}$ is close to 1. In eq.(5.3) the branch of the logarithm which yields the smallest $a\|B_\mu^\Lambda(x)\|$ should be taken.

In this expression the gauge field $U$ is considered to be a function of $B$ and $q$ according to eq.(5.2).

*5.2 Functional integral and vertex functions*

In the gauge fixed theory the dynamical variables are the quantum field $q$ and the ghost fields $\bar{c}$ and $c$. The a priori probability distribution for $q_\mu(x)$ derives from the SU($N$) invariant distribution $dU(x,\mu)$ of the corresponding link matrix $U(x,\mu)$. From eq.(5.2) and subsect. A.5 it is straightforward to show that

$$dU(x,\mu) = {\rm constant} \times \prod_{a=1}^{N^2-1} dq_\mu^a(x) \det\left\{J\left(g_0 a q_\mu(x)\right)\right\}. \tag{5.10}$$

In perturbation theory the jacobian appearing in this formula gives rise to a set of vertices, the "measure" vertices, which must be included in the Feynman rules (see ref.[6] for further details).

The partition function of the lattice theory is defined through

$$Z[B,J,\bar{\eta},\eta] = \frac{1}{\mathcal{N}} \int \prod_{x,\mu} dU(x,\mu) \prod_{y,a} d\bar{c}^a(y) \prod_{z,b} dc^b(z)$$
$$\times \exp\left\{-S_{\rm tot}[B,q,\bar{c},c] + (J,q) + (\bar{\eta},c) + (\bar{c},\eta)\right\}. \tag{5.11}$$

As in the continuum theory we have introduced source fields for the quantum fields. In general the situation is very much the same as in subsect. 2.2. Most comments made there carry over literally and this is also true for the definition of the free energy $W$ and the vertex functional $\Gamma$ if we set

$$\frac{\delta W}{\delta J_\mu^a(x)} = a^{-4} \frac{\partial W}{\partial J_\mu^a(x)} \tag{5.12}$$

(and similarly for the other functional derivatives).

The lattice vertex functions $\Gamma^{(j,k,l)}$ are obtained through the expansion

$$\Gamma[B,Q,\overline{C},C] = \sum_{j,k,l=0}^{\infty} \frac{1}{j!\,k!\,(l!)^2} \int_{-\pi/a}^{\pi/a} \frac{d^4 p_1}{(2\pi)^4} \cdots \frac{d^4 s_l}{(2\pi)^4}$$

$$\times (2\pi)^4 \delta_P(p_1 + \ldots + s_l) \, \Gamma^{(j,k,l)}(p_1,\ldots,s_l)^{a_1\ldots d_l}_{\mu_1\ldots\nu_k}$$

$$\times \tilde{B}^{a_1}_{\mu_1}(-p_1)\ldots\tilde{B}^{a_j}_{\mu_j}(-p_j)\tilde{Q}^{b_1}_{\nu_1}(-q_1)\ldots\tilde{Q}^{b_k}_{\nu_k}(-q_k)$$

$$\times \tilde{\overline{C}}^{c_1}(-r_1)\ldots\tilde{\overline{C}}^{c_l}(-r_l)\tilde{C}^{d_1}(-s_1)\ldots\tilde{C}^{d_l}(-s_l). \quad (5.13)$$

As opposed to the continuum theory, the momenta $p_1,\ldots,s_l$ are restricted to the Brillouin zone of the hypercubic lattice. $\delta_P$ denotes a periodic Dirac $\delta$–function,

$$\delta_P(q) = \sum_{n \in \mathbb{Z}^4} \delta(q - 2\pi n/a), \quad (5.14)$$

and our conventions on the Fourier transformation of lattice fields are described in appendix A. The relations between the vertex functions and the lattice correlation functions $G^{(j,k,l)}$ (which one defines by expanding the free energy) are the same as in the continuum theory. In particular, eqs.(2.15)–(2.17) remain valid.

## 6. BRS and shift symmetry on the lattice

The general discussion of the gauge fixing procedure in ref.[14] applies to the lattice theory considered in the present paper. In particular, the existence of an exact BRS symmetry is guaranteed by the abstract arguments given there. In this section we shall not rely on these results, but simply write down the transformation and verify algebraically that the theory is invariant. Using some of the identities established along the way we are then able to prove that there is an exact shift symmetry and that $W[B,0,0,0] = 0$.

## 6.1 Lattice BRS symmetry

The BRS variation $\delta_{\text{BRS}}F$ of any function $F[q,\bar{c},c]$ is defined through

$$\delta_{\text{BRS}}F = a^4 \sum_{x \in \Lambda} \left\{ \delta_{\text{BRS}}q_\mu^a \frac{\delta F}{\delta q_\mu^a} + \delta_{\text{BRS}}\bar{c}^a \frac{\delta F}{\delta \bar{c}^a} + \delta_{\text{BRS}}c^a \frac{\delta F}{\delta c^a} \right\}, \qquad (6.1)$$

where

$$\delta_{\text{BRS}}q_\mu = \left[ J(g_0 a q_\mu)^{-1} D_\mu + g_0 \text{Ad} q_\mu \right] c, \qquad (6.2)$$

$$\delta_{\text{BRS}}\bar{c} = \lambda_0 D_\mu^* q_\mu, \qquad (6.3)$$

$$\delta_{\text{BRS}}c = -g_0 c^2. \qquad (6.4)$$

A particularly interesting function to consider is the link variable $U(x,\mu)$ [cf. eq.(5.2)]. From the definition of the matrix $J$ given in appendix A and the equations above it is straightforward to show that

$$\delta_{\text{BRS}}U(x,\mu) = g_0 U(x,\mu) c(x + a\hat{\mu}) - g_0 c(x) U(x,\mu). \qquad (6.5)$$

The BRS transformation thus amounts to an infinitesimal gauge transformation when applied to the total field $U$. In particular, the Wilson action $S[U]$ and the a priori measure in the functional integral are invariant.

Another important property of the lattice BRS transformation is that

$$\delta_{\text{BRS}}(\delta_{\text{BRS}}q_\mu) = 0. \qquad (6.6)$$

To establish this identity we first use eqs.(6.4) and (6.5) to show that

$$\delta_{\text{BRS}}\left(\delta_{\text{BRS}}U(x,\mu)\right) = 0. \qquad (6.7)$$

Next we note that

$$\delta_{\text{BRS}}\left(\delta_{\text{BRS}}U(x,\mu)\right) =$$

$$\delta_{\text{BRS}}\left(\delta_{\text{BRS}}q_\mu^a(x)\right) \frac{\partial U(x,\mu)}{\partial q_\mu^a(x)} - \delta_{\text{BRS}}q_\mu^a(x) \delta_{\text{BRS}}q_\mu^b(x) \frac{\partial^2 U(x,\mu)}{\partial q_\mu^a(x) \partial q_\mu^b(x)} \qquad (6.8)$$

(no sum over $\mu$ is implied here). Because of the anti-commuting nature of $\delta_{\text{BRS}}q_\mu^a$ the last term in this equation vanishes. Taking eq.(6.7) into account, the desired result, eq.(6.6), follows.

We still need to prove that the total action is BRS invariant. To this end we rewrite the gauge fixing term and the ghost action in the form

$$S_{\text{gf}} = \frac{\lambda_0}{2} \left( D^*_\mu q_\mu, D^*_\nu q_\nu \right), \tag{6.9}$$

$$S_{\text{FP}} = - \left( \bar{c}, D^*_\mu \delta_{\text{BRS}} q_\mu \right). \tag{6.10}$$

Using eq.(6.6) it is then trivial to verify that

$$\delta_{\text{BRS}} S_{\text{gf}} + \delta_{\text{BRS}} S_{\text{FP}} = 0. \tag{6.11}$$

Taken together our results imply $\delta_{\text{BRS}} S_{\text{tot}} = 0$, which shows that the BRS transformation (6.1)–(6.4) is indeed a symmetry of the lattice theory.

We finally note that

$$\delta_{\text{BRS}}(\delta_{\text{BRS}} c) = 0. \tag{6.12}$$

Compared to the BRS symmetry in the continuum theory there is thus hardly any difference. All the identities that are important for the proof of the finiteness of the renormalized vertex functional hold exactly.

### 6.2 Shift symmetry

On the lattice the shift symmetry variation generates the transformations

$$e^{aB_\mu(x)} \to e^{g_0 a \omega_\mu(x)} e^{aB_\mu(x)}, \tag{6.13}$$

$$e^{g_0 a q_\mu(x)} \to e^{g_0 a q_\mu(x)} e^{-g_0 a \omega_\mu(x)}, \tag{6.14}$$

where $\omega^a_\mu(x)$ is an arbitrary vector field. From eq.(5.2) it is obvious that the total field $U$ (and hence the Wilson action $S[U]$) is invariant under such transformations.

To write down the shift symmetry variation $[\delta_s F]^a_\mu$ of any given function $F[B, q]$ explicitly and to study its properties, it is useful to introduce the related function

$$\hat{F}[V, U] = F[B, q], \qquad V(x, \mu) = e^{aB_\mu(x)}. \tag{6.15}$$

We then define

$$[\delta_s F]^a_\mu = \left. \frac{\delta}{\delta \omega^a_\mu} \hat{F} \left[ e^{g_0 a \omega} V, U \right] \right|_{\omega = 0}. \tag{6.16}$$

This is equivalent to

$$[\delta_{\rm s} F]^a_\mu = g_0 \left[J(aB_\mu)^{-1}\right]^{ab} \frac{\delta F}{\delta B^b_\mu} - \frac{\delta F}{\delta q^b_\mu} \left[J(g_0 a q_\mu)^{-1}\right]^{ba}, \quad (6.17)$$

as one may show using the formulae for the differential of the exponential mapping quoted in subsect. A.5.

A somewhat surprising property of the shift symmetry variation is that

$$[\delta_{\rm s}(\delta_{\rm BRS} F)]^a_\mu = \delta_{\rm BRS}([\delta_{\rm s} F]^a_\mu) \quad (6.18)$$

for any function $F$ which does not depend on the ghost fields. To show this we first observe that the BRS variation of such functions may be computed through

$$\delta_{\rm BRS} F = \left. \frac{\partial}{\partial t} \hat{F}[V, U_t] \right|_{t=0}, \quad (6.19)$$

where $U_t$ is the one-parameter curve of lattice gauge fields given by

$$U_t(x, \mu) = {\rm e}^{-t g_0 c(x)} U(x, \mu) \, {\rm e}^{t g_0 c(x+a\hat{\mu})} \quad (6.20)$$

[cf. eq.(6.5)]. Taking eq.(6.16) into account, we thus conclude that

$$[\delta_{\rm s}(\delta_{\rm BRS} F)]^a_\mu = \left. \frac{\delta}{\delta \omega^a_\mu} \frac{\partial}{\partial t} \hat{F}\left[{\rm e}^{g_0 a \omega} V, U_t\right] \right|_{\omega = t = 0}. \quad (6.21)$$

The validity of eq.(6.18) now follows, since the right hand side of the equation is obtained by interchanging the differentiations with respect to $\omega$ and $t$ in the expression above.

In the continuum theory the shift symmetry variation of the total action is equal to the BRS variation of some local composite field [cf. eq.(3.14)]. We now show that the same is true on the lattice. From the above we already know that $[\delta_{\rm s} S]^a_\mu = 0$. The shift symmetry variation of the gauge fixing term and the ghost action is given by

$$[\delta_{\rm s} S_{\rm gf}]^a_\mu = \lambda_0 \left(D^*_\nu q_\nu, [\delta_{\rm s} D^*_\rho q_\rho]^a_\mu\right), \quad (6.22)$$

$$[\delta_{\rm s} S_{\rm FP}]^a_\mu = -\left(\bar{c}, [\delta_{\rm s}(\delta_{\rm BRS} D^*_\rho q_\rho)]^a_\mu\right). \quad (6.23)$$

As a result we have

$$[\delta_{\rm s} S_{\rm tot}]^a_\mu = \delta_{\rm BRS}\left(\bar{c}, [\delta_{\rm s} D^*_\rho q_\rho]^a_\mu\right), \quad (6.24)$$

where use has been made of the identity (6.18) with $F[B,q] = D^*_\rho q_\rho$. Note that the scalar product in eq.(6.24) reduces to a local composite field, since the shift variation is a local operation.

We can now prove that $W[B,0,0,0]$ is independent of the background field $B$ by observing that the expectation value on the right hand side of the equation

$$g_0 \left[J(aB_\mu)^{-1}\right]^{ab} \frac{\delta}{\delta B^b_\mu} W[B,0,0,0] = -\langle[\delta_s S_{\text{tot}}]^a_\mu\rangle_B \qquad (6.25)$$

vanishes as a consequence of eq.(6.24) and the BRS invariance of the theory. In particular, the identity (3.16) is also valid on the lattice.

## 7. Renormalization of the lattice theory

At this point it should be rather obvious that our discussion of the renormalization of the effective action in sect. 4 carries over to the lattice theory. A few additional remarks may however be helpful.

(a) We again introduce the source terms (4.2) in the functional integral and define the renormalized vertex functional $\Gamma_R$ through eqs.(4.5)–(4.7) [the factor $\mu^\epsilon$ should be dropped of course]. The renormalization constants $Z_1, Z_3$ and $\tilde{Z}_3$ are formal powers series in $g$ with coefficients depending on the lattice spacing $a$ and the renormalized gauge parameter $\lambda$. We may, for example, adopt a minimal subtraction scheme, where the coefficients are polynomials in $\ln(a\mu)$ with no constant term. The mass $\mu$ then plays the same rôle as the normalization mass in the case of dimensional regularization.

(b) The continuum limit of the theory is taken in momentum space, i.e. we consider the renormalized vertex functions at non-exceptional momenta and send the lattice spacing $a$ to 0. For appropriately chosen renormalization constants, the existence of the limit in the absence of the background field has been established by Reisz [7–9].

(c) From the BRS invariance of the lattice theory one derives the identity

$$a^4 \sum_{x\in\Lambda} \left\{ \frac{\delta\Gamma_R}{\delta Q^a_\mu} \frac{\delta\Gamma_R}{\delta K^a_\mu} - \frac{\delta\Gamma_R}{\delta C^a} \frac{\delta\Gamma_R}{\delta L^a} + \lambda(D^*_\mu Q_\mu)^a \frac{\delta\Gamma_R}{\delta \overline{C}^a} \right\} = 0. \qquad (7.1)$$

The consequences of the background gauge invariance and the shift symmetry are as in the continuum theory.

(d) From power counting, the background gauge invariance and the discrete lattice symmetries one infers that a sufficiently general expression for the singular part $\Delta\Gamma_{R,n}$ of the vertex functional is given by

$$\Delta\Gamma_{R,n} = a^4 \sum_{x \in \Lambda} \left\{ c_1 \mathcal{O}_1(x) + c_2 \mathcal{O}_2(x) + c_3 \mathcal{O}_3(x) \right\}, \qquad (7.2)$$

where $\mathcal{O}_k(x)$, $k = 1, 2, 3$, are local lattice fields which reduce to the continuum fields appearing in eq.(4.16) in the limit $a \to 0$. Moreover we can insist that the coefficients $c_k$ are polynomials in $\ln(a\mu)$ with no constant term.

(e) Following the lines of subsect. 4.3, we can now establish the finiteness of the renormalized vertex functional in the continuum limit for non-zero background fields $B$. It is obvious that the symmetries already present on the lattice are preserved in the limit.

## 8. Concluding remarks

A final point we wish to make is that up to finite renormalizations the renormalized vertex functional is the same for any two regularization schemes, provided the BRS, background gauge and shift symmetries are respected in both cases. More precisely, in the limit where the cutoff has been removed, the corresponding vertex functionals $\Gamma_R$ and $\Gamma'_R$ are related by

$$\Gamma'_R[B, Q, \overline{C}, C, K, L; \mu, g, \lambda] =$$
$$\Gamma_R[B, z_3^{1/2} Q, \tilde{z}_3^{1/2} \overline{C}, \tilde{z}_3^{1/2} C, \tilde{z}_3^{1/2} K, z_3^{1/2} L; \mu, z_1 z_3^{-3/2} g, z_3^{-1} \lambda], \quad (8.1)$$

where $z_1$, $z_3$ and $\tilde{z}_3$ are formal power series in the renormalized coupling $g$ with coefficients depending on the renormalized gauge parameter $\lambda$. For simplicity the normalization mass $\mu$ has been taken to be the same in both schemes.

To prove this one proceeds essentially as in sect. 4. One first fixes the renormalization constants so that eq.(8.1) holds for $B = 0$. One then assumes that the equation is satisfied for non-zero $B$ at all loop orders $l < n$ and shows that the difference $\Delta\Gamma_{R,n}$ between the functionals at $n$–loop order must vanish as a result of the symmetry properties of $\Gamma_R$ and $\Gamma'_R$.

An important consequence of this remark is that one only needs to compute a set of propagator type diagrams to determine the renormalization constants $z_1$, $z_3$ and $\tilde{z}_3$ at any given order of perturbation theory. This is obvious for $z_3$ and $\tilde{z}_3$, but to calculate $z_1$ one usually needs to work out a 3–point vertex function. Since the background field propagator is inversely proportional to the coupling constant at tree-level of perturbation theory,

$$\Gamma_R^{(2,0,0)}(p,-p)_{\mu\nu}^{ab} = -\delta^{ab}(\delta_{\mu\nu}p^2 - p_\mu p_\nu)/g^2 + \ldots, \tag{8.2}$$

and since the background field is not renormalized, we may instead use $\Gamma_R^{(2,0,0)}$ to extract $z_1 z_3^{-3/2}$. In this way the number of diagrams that must be calculated is significantly reduced. Moreover it is usually much simpler to compute diagrams with two rather than three external legs. In our two-loop computation of the relation between the bare lattice coupling and the $\overline{\text{MS}}$ coupling we have thus decided to make use of this possibility (further details will be given in ref.[6]).

# Appendix A

*A.1 Indices*

Lorentz indices $\mu,\nu,\ldots$ normally run from 0 to 3. In the context of dimensional regularization they run up to $D-1$, the dimension of space. Since the metric is euclidean it does not matter in which position these indices appear. Color vectors in the fundamental representation of SU($N$) carry indices $\alpha,\beta,\ldots$ ranging from 1 to $N$, while for vectors in the adjoint representation, Latin indices $a,b,\ldots$ running from 1 to $N^2-1$ are employed. Repeated indices are automatically summed over unless stated otherwise.

## A.2 Gauge group

The Lie algebra $\mathrm{su}(N)$ of $\mathrm{SU}(N)$ can be identified with the space of complex $N \times N$ matrices $X_{\alpha\beta}$ which satisfy

$$X^\dagger = -X, \qquad \mathrm{tr}\{X\} = 0, \tag{A.1}$$

where $X^\dagger$ denotes the adjoint matrix of $X$ and $\mathrm{tr}\{X\} = X_{\alpha\alpha}$ is the trace of $X$. We may choose a basis $T^a, a = 1, 2, \ldots, N^2 - 1$, in this space such that

$$\mathrm{tr}\{T^a T^b\} = -\tfrac{1}{2}\delta^{ab}. \tag{A.2}$$

For $N = 2$, for example, the standard basis is

$$T^a = \frac{\tau^a}{2i}, \quad a = 1, 2, 3, \tag{A.3}$$

where $\tau^a$ denote the Pauli matrices. With these conventions the structure constants $f^{abc}$, defined through

$$[T^a, T^b] = f^{abc} T^c, \tag{A.4}$$

are real and totally anti-symmetric under permutations of the indices.

The representation space of the adjoint representation of $\mathrm{su}(N)$ is the Lie algebra itself, i.e. the elements $X$ of $\mathrm{su}(N)$ are represented by linear transformations

$$\mathrm{Ad}\,X : \mathrm{su}(N) \mapsto \mathrm{su}(N). \tag{A.5}$$

Explicitly, $\mathrm{Ad}\,X$ is defined through

$$\mathrm{Ad}\,X \cdot Y = [X, Y] \quad \text{for all} \quad Y \in \mathrm{su}(N). \tag{A.6}$$

With respect to a basis $T^a$ the associated matrix $(\mathrm{Ad}\,X)^{ab}$ representing the transformation is given by

$$\mathrm{Ad}\,X \cdot T^b = T^a (\mathrm{Ad}\,X)^{ab}, \tag{A.7}$$

which is equivalent to

$$(\mathrm{Ad}\,X)^{ab} = -f^{abc} X^c, \qquad (\mathrm{Ad}\,X \cdot Y)^a = f^{abc} X^b Y^c, \tag{A.8}$$

in terms of the structure constants.

*A.3 Fields (continuum theory)*

An SU($N$) gauge potential is a vector field $A_\mu(x)$ with values in the Lie algebra su($N$). It may thus be written as

$$A_\mu(x) = A_\mu^a(x) T^a \tag{A.9}$$

with real components $A_\mu^a(x)$. Most other fields occurring in this paper are also Lie algebra valued. In particular, this is so for the field strength tensor

$$F_{\mu\nu} = \partial_\mu A_\nu - \partial_\nu A_\mu + [A_\mu, A_\nu], \tag{A.10}$$

the ghost fields and the source fields encountered in various places. The components of the ghost fields and the corresponding source fields take values in a Grassmann algebra.

For any two fields of the same type a scalar product is defined in a natural way. In the case of vector fields, for example, it is given by

$$(J, q) = \int d^D x \, J_\mu^a(x) q_\mu^a(x). \tag{A.11}$$

Note that the scalar product of bosonic fields is real and symmetric, while for fields with odd ghost number it is anti-symmetric.

The Fourier transform $\tilde{B}_\mu^a(p)$ of the background field is defined by

$$\tilde{B}_\mu^a(p) = \int d^D x \, e^{-ipx} B_\mu^a(x). \tag{A.12}$$

The same conventions apply to all other source fields.

*A.4 Lattice notations*

The lattice theory discussed in this paper lives on a hypercubic lattice with points $x = a(n_0, n_1, n_2, n_3)$, where the coordinates $n_\mu$ range over the set of integers. The unit vector in direction $\mu$ is denoted by $\hat{\mu}$.

A lattice gauge field is a field of SU($N$) matrices $U(x, \mu)$, where $x$ runs over all lattice points and $\mu = 0, \ldots, 3$. Under a gauge transformation $\Lambda(x)$ (which is an assignment of an SU($N$) matrix to each lattice point $x$), the gauge field transforms according to

$$U(x, \mu) \to \Lambda(x) U(x, \mu) \Lambda(x + a\hat{\mu})^{-1}. \tag{A.13}$$

As a substitute for the field strength tensor we introduce the plaquette field

$$P(x,\mu,\nu) = U(x,\mu)U(x+a\hat{\mu},\nu)U(x+a\hat{\nu},\mu)^{-1}U(x,\nu)^{-1}. \qquad (A.14)$$

The ghost fields $\bar{c}(x)$ and $c(x)$ are defined on the lattice points $x$ and are otherwise as in the continuum theory. The same comment applies to the "quantum" field $q_\mu(x)$ and all source fields including the background field $B_\mu(x)$.

On the lattice the scalar product (A.11) is replaced by

$$(J,q) = a^4 \sum_{x\in\Lambda} J^a_\mu(x) q^a_\mu(x). \qquad (A.15)$$

Scalar products of other lattice fields are defined analogously.

The Fourier transform of lattice vector fields involves an extra phase factor which is included to simplify the expressions for the propagators and vertices in perturbation theory. In the case of the background field we have

$$\tilde{B}^a_\mu(p) = a^4 \sum_{x\in\Lambda} e^{-ip(x+a\hat{\mu}/2)} B^a_\mu(x), \qquad (A.16)$$

and the same conventions apply to all other vector fields. The inverse of the transformation (A.16) is

$$B^a_\mu(x) = \int_{-\pi/a}^{\pi/a} \frac{d^4p}{(2\pi)^4} e^{ip(x+a\hat{\mu}/2)} \tilde{B}^a_\mu(p). \qquad (A.17)$$

For scalar fields, such as the ghost field $C^a(x)$, the standard formula

$$\tilde{C}^a(p) = a^4 \sum_{x\in\Lambda} e^{-ipx} C^a(x) \qquad (A.18)$$

is employed.

*A.5 Differential of the exponential mapping*

Let $X$ be an element of the Lie algebra $su(N)$. We then define a linear mapping $J(X) : su(N) \mapsto su(N)$ through

$$J(X) \cdot Y = e^{-X} \frac{d}{dt} e^{X+tY} \bigg|_{t=0} \qquad \text{for all } Y \in su(N). \qquad (A.19)$$

$J(X)$ is referred to as the differential of the exponential mapping. It is possible to show that

$$J(X) = 1 + \sum_{k=1}^{\infty} \frac{(-1)^k}{(k+1)!}(\mathrm{Ad}X)^k, \qquad (A.20)$$

which may symbolically be written as

$$J(X) = \frac{1 - \mathrm{e}^{-\mathrm{Ad}X}}{\mathrm{Ad}X}. \qquad (A.21)$$

With respect to a basis $T^a$ the associated matrix $J(X)^{ab}$ representing the transformation is given by

$$J(X) \cdot T^b = T^a J(X)^{ab}. \qquad (A.22)$$